\documentclass[final,3p,times,sort&compress,twocolumn]{elsarticle}

\usepackage{amssymb,amsmath}

\begin{document}

\title{Arnold cat map, Ulam method and time reversal}

\author[LPT]{L.~Ermann}
\ead{ermann@irsamc.ups-tlse.fr}

\author[LPT]{D.L.~Shepelyansky}
\address[LPT]{\mbox{Laboratoire de Physique Th\'eorique du CNRS, IRSAMC, 
Universit\'e de Toulouse, UPS, F-31062 Toulouse, France}}
\ead[url]{http://www.quantware.ups-tlse.fr/dima}


\begin{abstract}
We study the properties of the Arnold cap map
on a torus with a several periodic sections
using the Ulam method. This approach generates
a Markov chain with the Ulam matrix approximant.
We study numerically the spectrum and eigenstates of this matrix
showing their relation with the Fokker-Plank
relaxation and the Kolmogorov-Sinai entropy.
We show that, in the frame of the Ulam method, 
the time reversal property of the map
is preserved only on a short Ulam time
which grows only logarithmically with the matrix size. 
Parallels with  the evolution in a regime of quantum chaos are
also discussed.
\end{abstract}

\maketitle

\section{Introduction}
The Arnold cat map \cite{arnold} 
is the cornerstone model of classical dynamical chaos 
\cite{sinai,chirikov,lichtenberg}. 
This symplectic map belongs to the class of Anosov systems,
it has the positive Kolmogorov-Sinai entropy
$h\approx 0.96$
and is fully chaotic \cite{lichtenberg}. The map has the form
\begin{equation}
\label{eq1}
\bar{p}=p+x \; \mbox{(mod} \;\mbox{L)}\;\;, \;\; 
\bar{x}=x+\bar{p} \;\mbox{(mod} 
\;\mbox{1)}\;.
\end{equation}
Here the first equation can be seen as a kick which changes the momentum 
$p$ of a particle on a torus 
while the second one corresponds to a free phase rotation
in the interval $-0.5\leq x < 0.5$; bars mark the new values of 
canonical variables $(x,p)$.
The map dynamics takes place on a torus of integer length $L$ in the 
$p$ direction with $-L/2 < p \leq L/2$. 
The usual case of the Arnold cap map
corresponds to $L=1$ but it is also possible to study the chaotic
properties of the map on a torus of longer integer size 
$L>1$ as it has been discussed e.g. in \cite{demon}.
For $L \gg 1$ the spreading in $p$
is characterized by a diffusive process described by the 
Fokker-Planck  equation:
\begin{equation}
\label{eq2}
 \partial w(p,t)/ \partial t = D/2\; \; 
\partial^2 w(p,t)/ \partial^2 p,
\end{equation}
where the diffusion coefficient $D \approx <x^2>= 1/12$
$w(p,t)$ is a probability distribution over momentum
and $t$ being integer time measured in number of iterations.
As a result for times
$t \gg L^2/D$ the distribution converges to the ergodic equilibrium
with a homogeneous density in the plane $(x,p)$.
The exponential  convergence to the equilibrium state
is determined by the second
eigenvalue $\lambda_2$ 
of evolution (\ref{eq2}) on one map iteration
with  $|\lambda|=\exp(-\Gamma_D) < 1$
and the convergence rate 
\begin{equation}
\label{eq3}
 \Gamma_D=2\pi^2 D/L^2 \approx 1.6449/L^2 \;\; ;
\end{equation}
the fist eigenvalue is $\lambda_1=1$.

The dynamical equations (\ref{eq1}) 
are reversible in time, e.g. 
at the middle of free rotation,
but, due to chaos and exponential instability of motion,
small round-off errors break time reversal
leading to an irreversible relaxation to
the ergodic equilibrium \cite{demon}.

In this work we investigate the transition from
dynamical behavior to statistical description
using the Ulam method proposed in 1960 \cite{ulam}.
According to this method the whole phase space is
covered by equidistant lattice
($N=N_p \times N_x$ in our case). Then the transition probabilities
from cell to cell are determined by
propagating a large number of trajectories
$N_{tr}$ from one initial cell $j$ to all
other cells $i$ after one iteration of the map
(we used here $N_{tr}=10^5$). In this way
we generate the Markov chain \cite{markov}
with a transition matrix
$S_{ij}=N_{ij}/N_{tr}$, where $N_{ij}$ is the number
of trajectories arrived from cell $j$ to cell $i$.
By construction we have $\sum_{i=1}^{N}S_{ij}=1$
and thus the matrix $S$ belongs to the class of Perron-Frobenius operators
\cite{arnold,sinai,mbrin}. It is proven that for hyperbolic maps 
in one and higher dimensions
the Ulam method converges to the spectrum of continuous system 
\cite{li,liverani,froyland}.
At the same time it is known that in certain cases the Ulam method
gives significant modifications of the spectrum compared to
the case of the continuous Perron-Frobenius operators \cite{liverani}.
Indeed, for Hamiltonian maps with divided phase space
the spectrum is completely modified
(see discussions in \cite{zhirovattr,frahmstmap})
due to penetration of trajectories inside stability
islands. From a physical view point 
the discretization corresponds to an effective noise
in canonical variables which amplitude is
equal to the cell size. Since an arbitrary 
small noise gives propagation of trajectories
inside stability islands \cite{lichtenberg}
the spectrum of the Ulam matrix approximant of size $N$
in such a case differs from its continuous limit.
A generalization of the Ulam method, based on one ergodic
trajectory, allows to obtain a convergent spectrum
for dynamics on a chaotic component \cite{frahmstmap}.
\begin{figure}
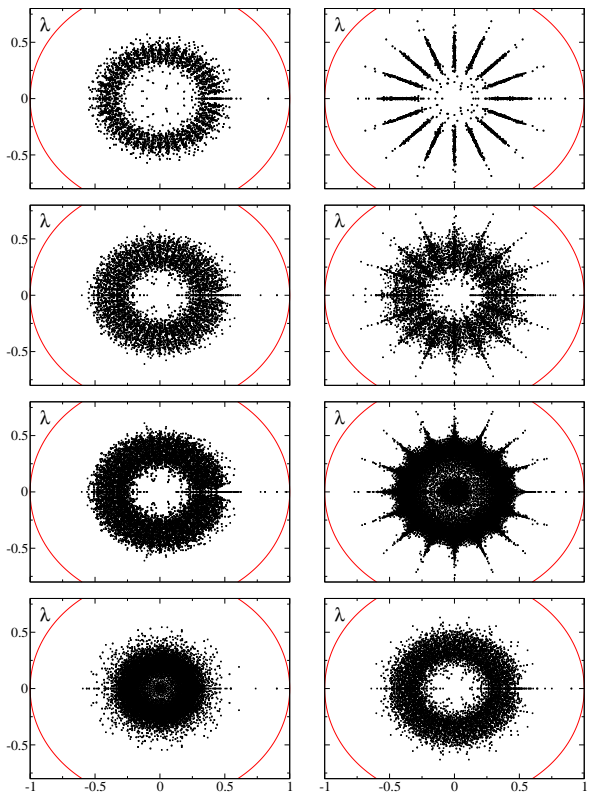

\begin{center}
\includegraphics[width=0.23\textwidth]{fig1a.eps}
\includegraphics[width=0.23\textwidth]{fig1b.eps}\\   
\includegraphics[width=0.23\textwidth]{fig1c.eps}
\includegraphics[width=0.23\textwidth]{fig1d.eps}\\
\includegraphics[width=0.23\textwidth]{fig1e.eps}
\includegraphics[width=0.23\textwidth]{fig1f.eps}\\
\includegraphics[width=0.23\textwidth]{fig1g.eps}
\includegraphics[width=0.23\textwidth]{fig1h.eps}
\caption{Complex spectrum $\lambda$
of the Ulam matrix approximant $S$ for the Arnold cat 
map. Top three rows: right column has $N_x=47$
and $L=3,4,8$ (from top to bottom);
left column has $N_x=43$
and $L=3,4,8$ (from top to bottom).
Bottom row is for $L=4$ with
$N_x=50$(left)
and $N_x=51$ (right).
The total matrix size is $N=N_p N_x$
with $N_p=L N_x$
and all eigenvalues are shown for each panel.
Unit circle is shown in red. 
}\label{fig1}
\end{center}
\end{figure}

The majority of numerical studies
with the Ulam method has been done for one-dimensional maps
(see e.g. \cite{tel,froyland1,ermann}) but recently
the studies were extended to the two-dimensional maps
(see e.g. \cite{zhirovattr,frahmstmap,froyland2,ermannweyl}).
In a certain respect the interest to such studies
was generated by similarities between properties of 
the Ulam matrix approximant for dynamical maps,
which can be viewed as the Ulam networks,
and the Google matrix of the World Wide Web
as it is discussed in \cite{zhirovattr,ermann}.
For 2D dissipative maps it was found that
the spectrum is characterized by the fractal Weyl law
\cite{zhirovattr,ermannweyl}.

In a difference from the previous studies of the Ulam method
in 2D maps here we choose the Arnold cat map
on a torus of size $L$ since it is fully chaotic,
it has well defined 
diffusive relaxation to the ergodic state at large $L$,
and it is time reversible. Thus the aim of this work is
to understand the interplay of all these features 
in the frame of the Ulam method and the 
finite size Markov chain 
with the Ulam matrix approximant
$S$ generated by this method.

The paper is composed as follows:
in Section 2 we describe the 
properties of spectrum and eigenstates of the matrix $S$,
the features of time reversal are analyzed in
Section 3 and discussion of the results
is presented in Section 4.

\section{Spectrum and eigenstates of 
the Ulam matrix approximant}

The complex eigenvalues $\lambda_i$
and right eigenvectors $\psi_i$ 
of the Ulam matrix approximant ${\bf S}$
satisfy the equation
${\bf S} \psi_i = \lambda_i \psi_i$
and are determined numerically by
direct dioganalization. In agreement with the
Perron-Frobenius theorem \cite{mbrin}
the maximal eigenvalue is $\lambda_1=1$
with the corresponding eigenstate being 
real, nonnegative and homogeneously distributed
over the whole phase space.

The global distributions
of eigenvalues $\lambda_i$ in the complex plane
are shown in Fig.~\ref{fig1} for even and odd
number of cells. Usually we keep
$N_p=L N_x$ to have exactly the same
amount of cells in each of $L$ sections of the continuous
map. The results show that 2D distributions
are different for even and prime values
of $N_x$ (see Fig.~\ref{fig1}).
For the even case $\lambda$-values 
are homogeneous inside a circle of
a certain radius. For 
the odd case the distribution
has a form of a ring without 
eigenvalues at $|\lambda| \approx 0$
(or with a smaller density at zero).
The arithmetic properties of 
the number of cells $N_x$ and $N_p=L N_x$
play a visible role. Thus for
$N_x=47$ we have a 
formation of star with 16 star rays
while for $N_x=43$ there are 
44 rays which are much less visible
(for $N_x=37$ we obtain a similar type of
distribution with 38 rays).
We obtain a similar type of ring spectrum also for
$N_x=51$ (bottom right panel in Fig.~\ref{fig1}).
In the case when both
$N_x$ and $N_p$ are primes,
e.g. $N_x=47$, $N_p=191$,
(and hence we have only approximate
relation $N_p \approx L N_x$)
the visibility of rays
also decreases (data not shown).
\begin{figure}
\includegraphics[width=0.44\textwidth]{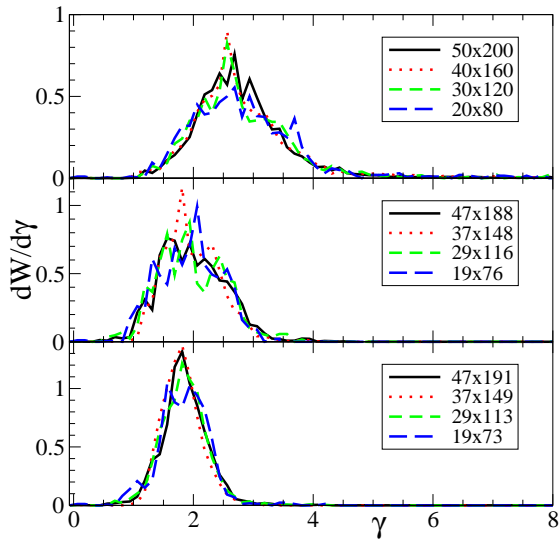}
\caption{(color online) Dependence of density of states 
$dW/d\gamma$ on the decay 
rate $\gamma$ for the Ulam matrix approximant
of the cat map with $L=4$. 
The number of cells $N_x$ and $N_p$ are even-even on 
top panel, prime-even on middle panel, and prime-prime on bottom panel. 
The corresponding values 
are shown on legends with the notation $N_x\times N_p$.
The densities of states are normalized by the condition 
$\int_0^{10}dW/d\gamma d\gamma=1$.
}\label{fig2}
\end{figure}

We expect that in the limit of 
large $N_x$ and $N_p=L N_x$
with fixed $L$ the distribution will
converge to a limiting one
in agreement with the
spirit of mathematical theorems about
the convergence of Ulam matrix approximant
for fully chaotic maps
\cite{li,liverani,froyland}.
A confirmation of this is seen in Fig.~\ref{fig2}
where the density distributions of eigenvalues
$d W/d\gamma$ are shown as a function of the relaxation rate
$\gamma=-2\ln |\lambda|$. Indeed, the density 
is essentially size independent showing two distinct distributions
for even and odd values of $N_x$. We suppose that
this difference between two cases
can be related to the effect of discretization 
on the continuous map symmetry $x \rightarrow -x$.
The third type of size independent distribution
appears in the case of prime values
of $N_x$ and $N_p \approx L N_x$
(see Fig.~\ref{fig2})
but in this case the difference should
be attributed to the fact
that this discretization does not
preserve exactly  $L$ identical classical
segments of the continuous map.
\begin{figure}
\includegraphics[width=0.44\textwidth]{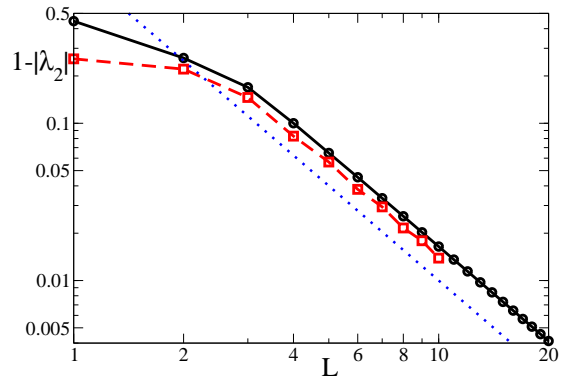}
\caption{(color online)
Spectral gap $\Delta=1-\vert\lambda_2\vert$ 
as a function of $L$.
Solid (black) and dashed (red) curves represent 
the cases of $N_x=23$ and $N_x=47$ 
respectively; $N_p=L N_x$. Dotted (blue) line shows 
the theoretical 
dependence $\Delta \propto L^{-2}$. 
}\label{fig3}
\end{figure}

The maximum of the distribution $d W/d\gamma$
is located approximately at
$\gamma=2 \approx 2 h$ corresponding to
the value of the Kolmogorov-Sinai entropy $h$.
Thus these $\gamma$ values describe the process
of exponential divergence of nearby trajectories
and are related to the exponential correlations
decay generated by chaotic dynamics.
In addition to these values $\gamma \sim 1$
there is also the value of $\lambda_2=\exp(-\gamma_2/2)$
which is positive and is
very close to the unit value $\lambda_1=1$.
It corresponds to the second eigenvalue of the Fokker-Plank
equation describing diffusive relaxation to the ergodic steady-state.
Indeed, the dependence  of the gap $\Delta = 1-\lambda_2$,
shown in Fig.~\ref{fig3}, is well in agreement with 
the dependence (\ref{eq3}):
a formal fit for $3 \leq L$ gives 
$\Delta \propto 1/L^{\mu}$ with the exponent
$\mu = 1.97$ for $N_x=23$ and
$1.94$ for $N_x=47$ being close to the theoretical value $\mu=2$.
The fit at the fixed exponent $\mu=2$
gives the numerical constant $a$ in the relation
$\Delta=a/L^2$ being $1.56 \pm 0.05$ for $N_x=23$
and $1.33 \pm 0.05$ for $N_x=47$ that is close to
the theoretical value $a=2\pi^2 D = 1.64...$. A small deviation
can be attributed
to the effect of finite size discretization.

\begin{figure}
\begin{center}
\includegraphics[height=0.44\textwidth,angle=-90]{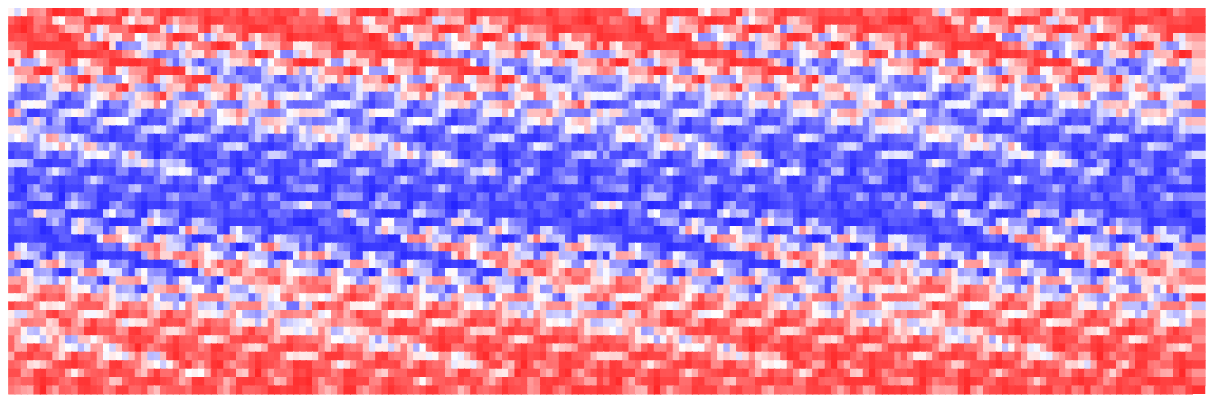}\\
\includegraphics[height=0.44\textwidth,angle=-90]{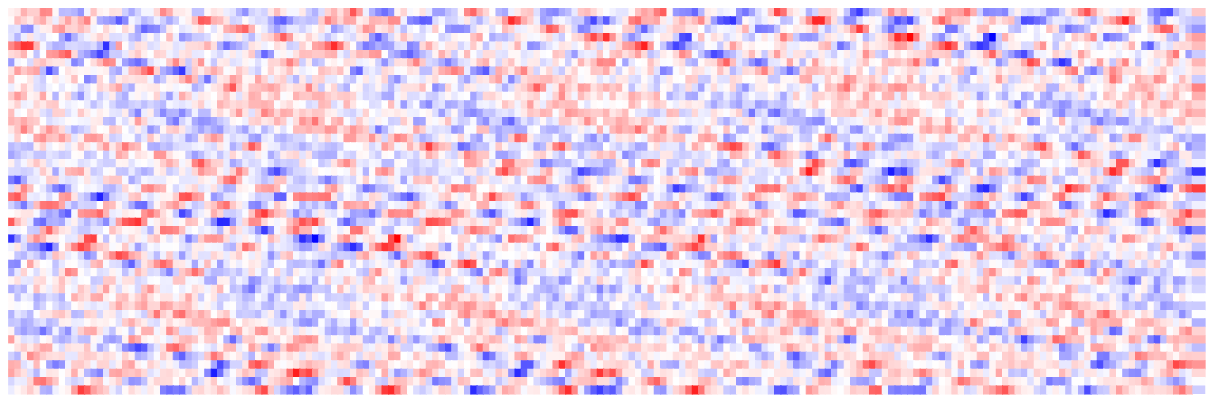}\\
\includegraphics[height=0.44\textwidth,angle=-90]{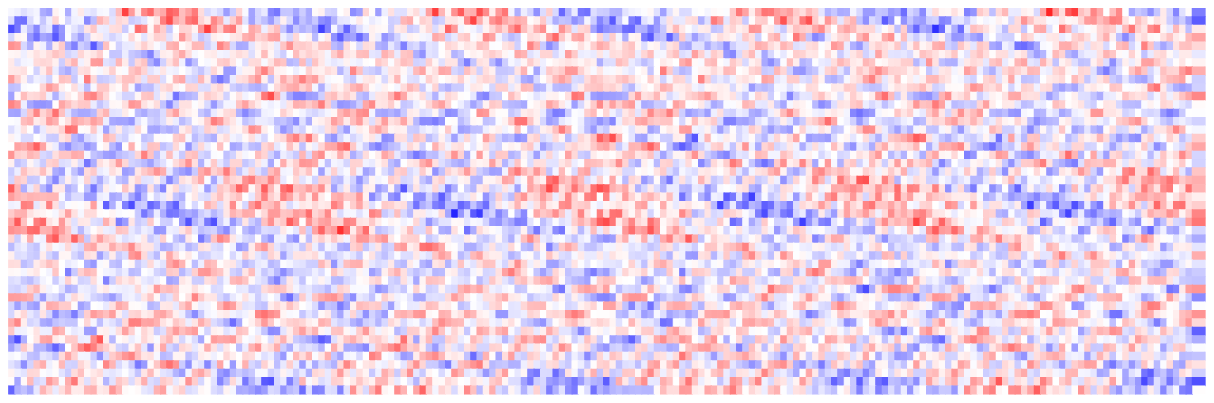}\\
\caption{(color online) 
Eigenstates of the matrix $S$ shown on
the whole phase space for $L=4$, $N_x=47$ and $N_p=LN_x$.
Top panel shows  the eigenfunction $\psi_2$, which is real and has
$\lambda_2=0.917$;
middle and bottom panels represent 
respectively real and imaginary part 
of the eigenstate $\psi_{27}$  
corresponding to the eigenvalue $\lambda_{27}\simeq0.516+i0.507$.
In all panels   red corresponds to positive, and blue to negative
values.
The phase space is rotated on
$90$ degrees clockwise 
so that positive values of $p$ are on  the right hand part
of the plot.
}\label{fig4}
\end{center}
\end{figure}

Two examples of eigenstates $\psi_2$ and $\psi_{27}$
of the matrix $S$ are shown in Fig.~\ref{fig4}
(the numbering is done in a decreasing order of $|\lambda|$).
According to the Fokker-Planck equation (\ref{eq2})
we expect to have two double degenerate
values of $\lambda_2$ with the corresponding
running wave eigenstates
$\psi_2 \propto \exp(\pm i k 2\pi p/L)$
with $k=\pm 1$ or their liner combination.
The numerically found eigenstate $\psi_2$
is real up to
a numerical level of precision of
matrix diagonalization, corresponding to a real $\lambda_2$
value. It shows a certain amplitude oscillations
along $p$ but its main feature is the sign
change along $x$, which is absent in the 
equation (\ref{eq2}). A dependence of eigenstates
on $x$ variable remains strongly visible 
and for other eigenstates (see  Fig.~\ref{fig4}).
For example, the state $\psi_{27}$ has density concentration
at points $x=0$ and $x=\pm 0.5$
corresponding to zero force point and
discontinuity point respectively.

These results show that the Fokker-Planck equation
gives only a fist approximation for
the statistical description of dynamics 
of the Arnold cat map. In contrast to that the Ulam
matrix approximant gives much more detailed
description. The understanding of all features of 
this statistical description 
requires further more detailed studies.
We note that according to (\ref{eq2})
there should be a series of eigenstates
with $\lambda_k=\exp(-\Gamma_D k^2)$.
We well resolved the case of $k=0, \pm 1$
but we found difficult 
to define accurately the corresponding higher values.
It is possible that they enter rapidly 
in the dense balk region of ring with
many eigenvalues and become mixed with them.
Probably larger values of $L$
should be studied to resolve such
eigenvalues in a better way.
We note that such a series of eigenvalues
has been seen for the Chirikov standard map
at the critical value of chaos parameter
where the diffusion rate is relatively small
and thus these eigenvalues are better 
separated from the balk region
\cite{frahmstmap}.

\section{Time reversal features and the Ulam time}

\begin{figure}
\includegraphics[width=0.44\textwidth]{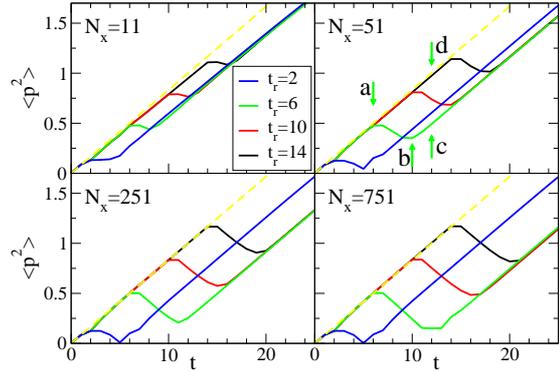}
\caption{(color online) Time evolution of $\langle p^2\rangle$ 
for an initial distribution with $p=0$ 
obtained with the Ulam matrix approximant of 
the Arnold cat map at $L=8$, $N_p=LN_x$
and $N_x=11$ (top-left panel), $51$ (top-right panel),
$251$ (bottom-left panel) and $751$ (bottom-right panels).
The time inversion is done at $t_r=2,6,10,14$.  
Yellow dashed line shows the 
theoretical prediction with $\langle p^2\rangle = D t$ 
and $D=1/12$.
The arrows in top-left panel shows the values of 
time $a,b,c,d$ plotted in Fig.~\ref{fig6}.
}\label{fig5}
\end{figure}

Even if the exact dynamics is time reversible
it becomes easily broken by small errors
due to dynamical chaos and exponential instability of motion
(see e.g. \cite{demon,qascat}).
In the quantum case the evolution is described 
by the linear  Schr\"odinger equation that together with the
uncertainty principle 
leads to stability of time reversal in respect to 
small errors \cite{demon,qascat,dls1983}.
Let's  now study the time reversal in the frame
of the Ulam method where the evolution is
described by a linear matrix transformation. 
For that we start from an initial
line in the phase space 
at $p=0$
with a homogeneous density in $x$, with
$w(p,t=0)=1/N_x$  and zero otherwise. 
Then we follow the evolution 
given by the matrix multiplication
$w(p,x,t+1)={\mathbf S} w(p,x,t)$
where time $t$ is  measured in 
the number of map iterations $t$.
The growth of the  second moment   $\langle p^2(t)\rangle$
as a function of time 
is shown in Fig.~\ref{fig5}.
The second moment grows diffusely with time
in agreement with the Fokker-Planck equation (\ref{eq2}). 
After $t_r$ iterations we perform time
reversal by inverting all momenta $p \rightarrow -p$.
We see that after $t_r$ the second moment
$\langle p^2(t)\rangle$ starts to decrease during
a certain time interval $t_{U}$,
where its value becomes minimal,
and after that 
the diffusion restarts again.
This time $t_{U}$ is the Ulam time scale during which 
we have anti-diffusive process which also describes relaxation
in a vicinity of big fluctuations
\cite{chirikovzhirov}. 

\begin{figure}
\includegraphics[width=0.22\textwidth]{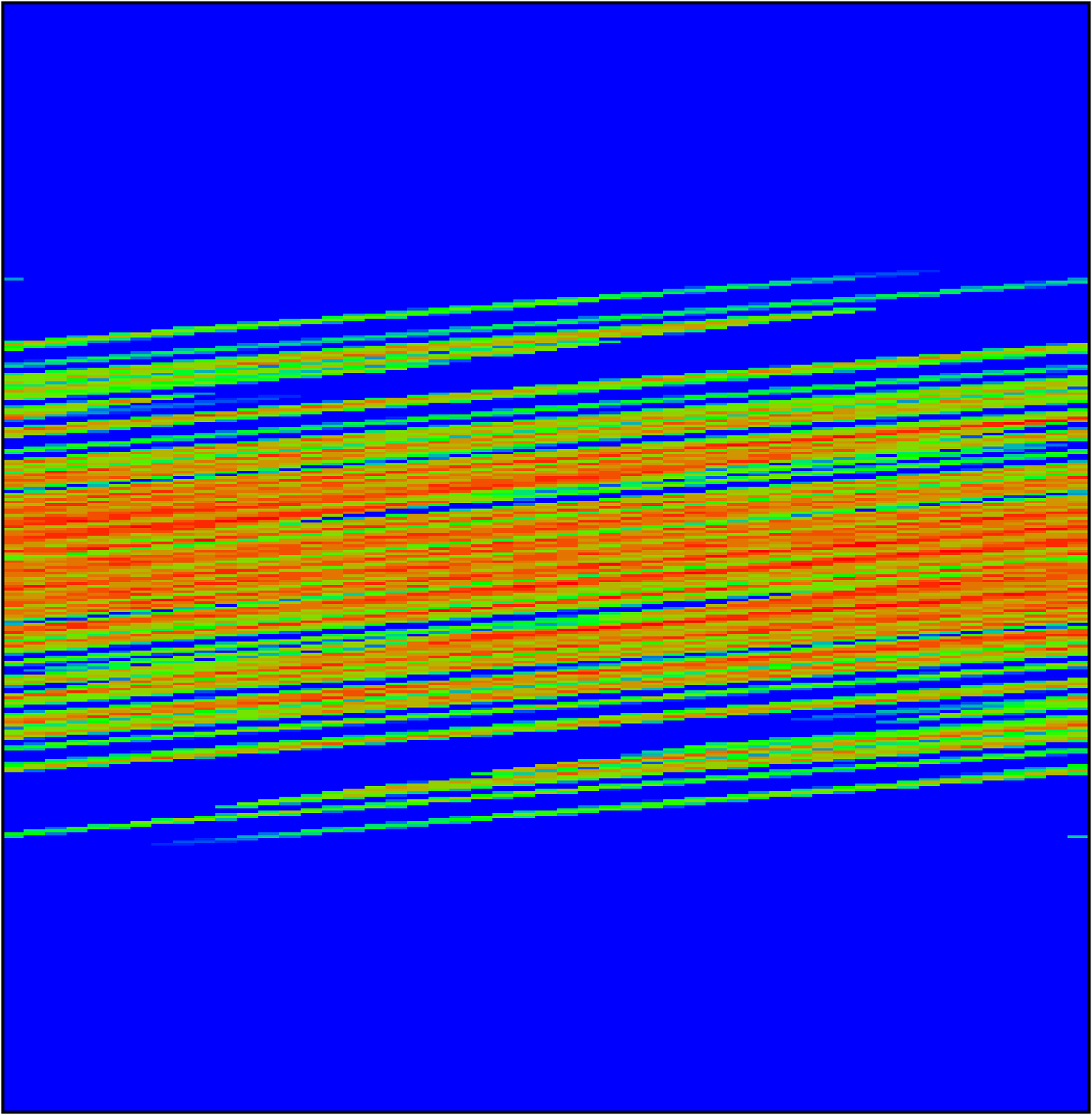}
\includegraphics[width=0.22\textwidth]{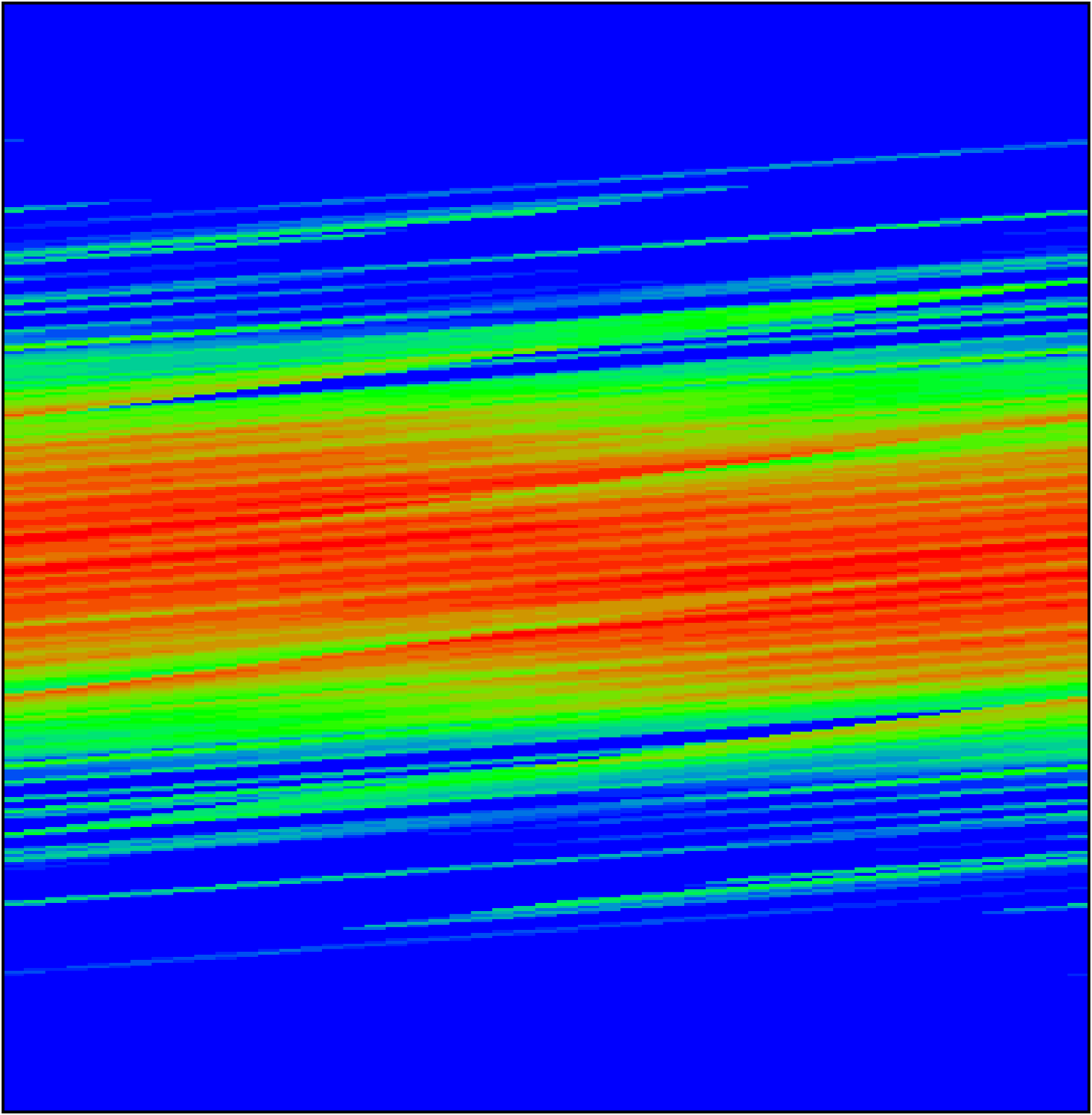}\\
\includegraphics[width=0.22\textwidth]{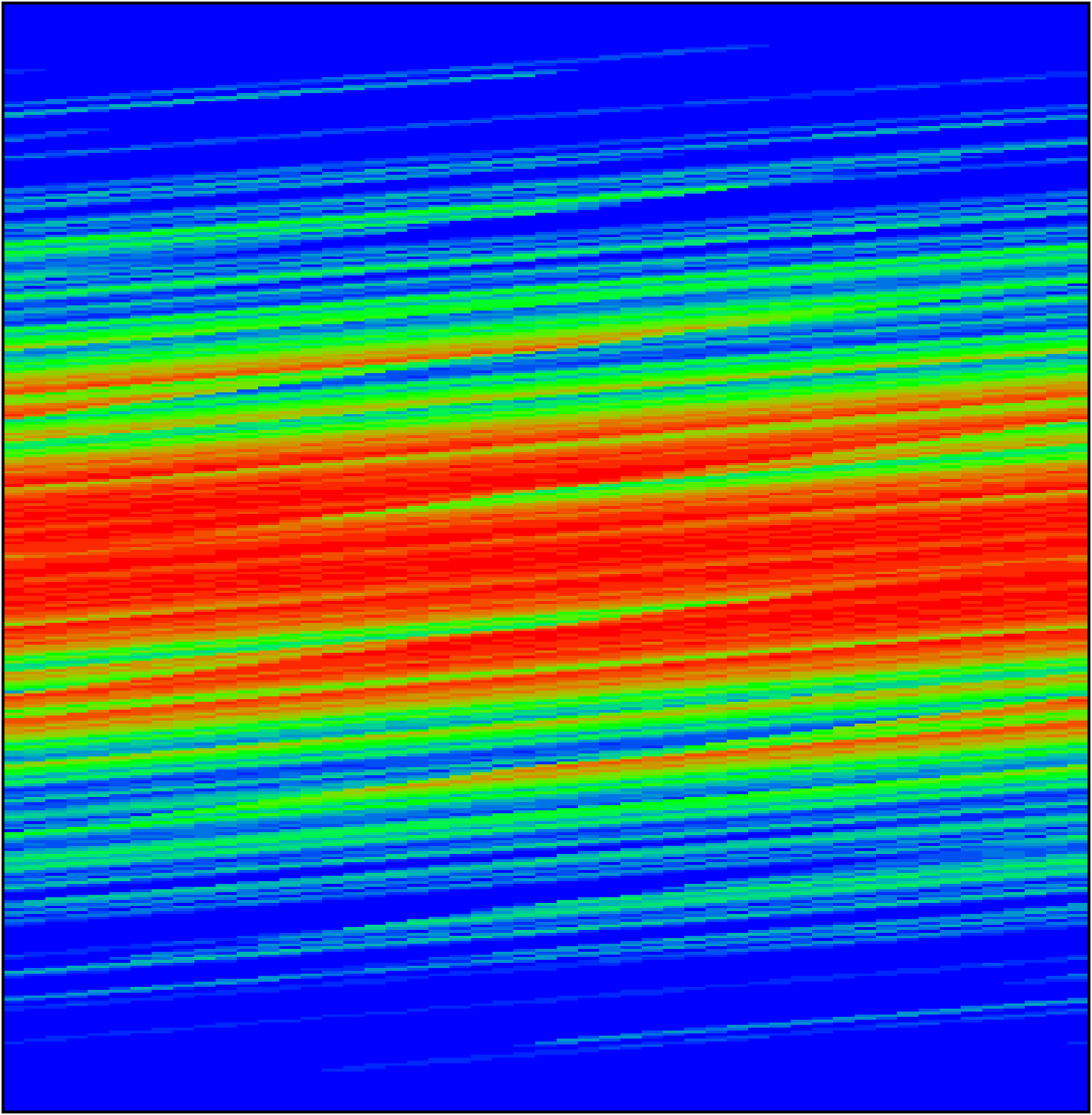}
\includegraphics[width=0.22\textwidth]{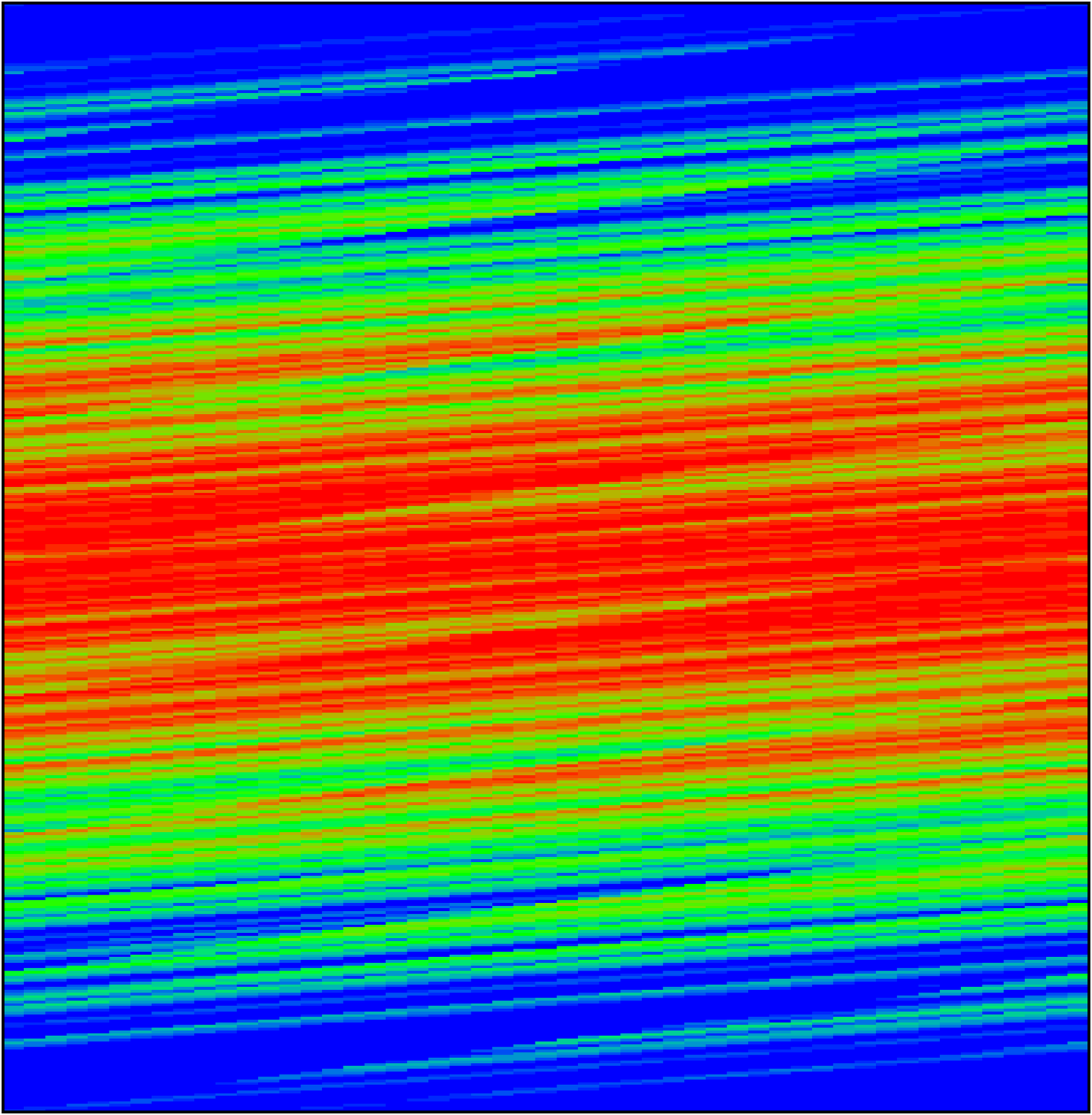}\\
\caption{(color online) Time evolution of a $p=0$ state with 
the Ulam matrix of 
the Arnold cat map at $L=8$, $N_x=51$ and $N_p=LN_x=408$.
Panels show distribution in the whole phase space
at the moments of time $a,b,c,d$ marked by arrows in Fig.~\ref{fig5}:
$(a)$ at $t=6$ (top-left panel); 
$(b)$ at  $t=10$ with $p$-inversion 
in $t=6$ (top-right panel); 
$(c)$: $t=12$ with $p$-inversion 
in $t=6$ (bottom-left panel);
$(d)$ at $t=12$ without $p$-inversion (bottom-right panel). 
Probability density is shown by color with
blue for zero density and red for maximal density
on a given panel.
The values of $\langle p^2\rangle$ are marked in green arrows on 
top-right panel of Fig. \ref{fig5}
}\label{fig6}
\end{figure}

The spreading of probability in the whole
phase space is displayed in Fig.~\ref{fig6}
at different moments of time. These data 
show that time reversal gives a temporary shrinking of the 
distribution followed by diffusive spreading continued.
Thus the time reversal is preserved
only on the Ulam time scale $t_U$.

\begin{figure}
\includegraphics[width=0.22\textwidth]{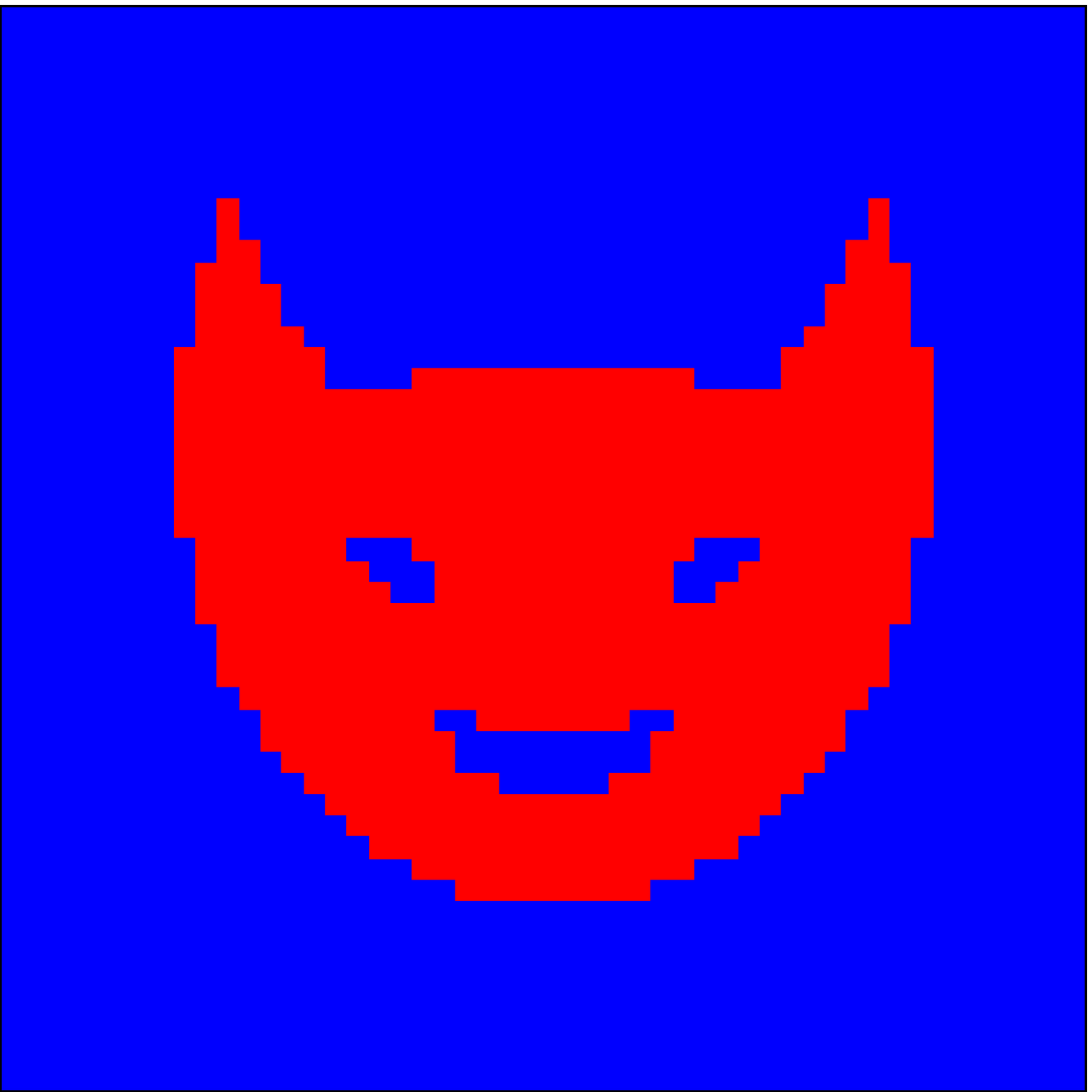}
\includegraphics[width=0.22\textwidth]{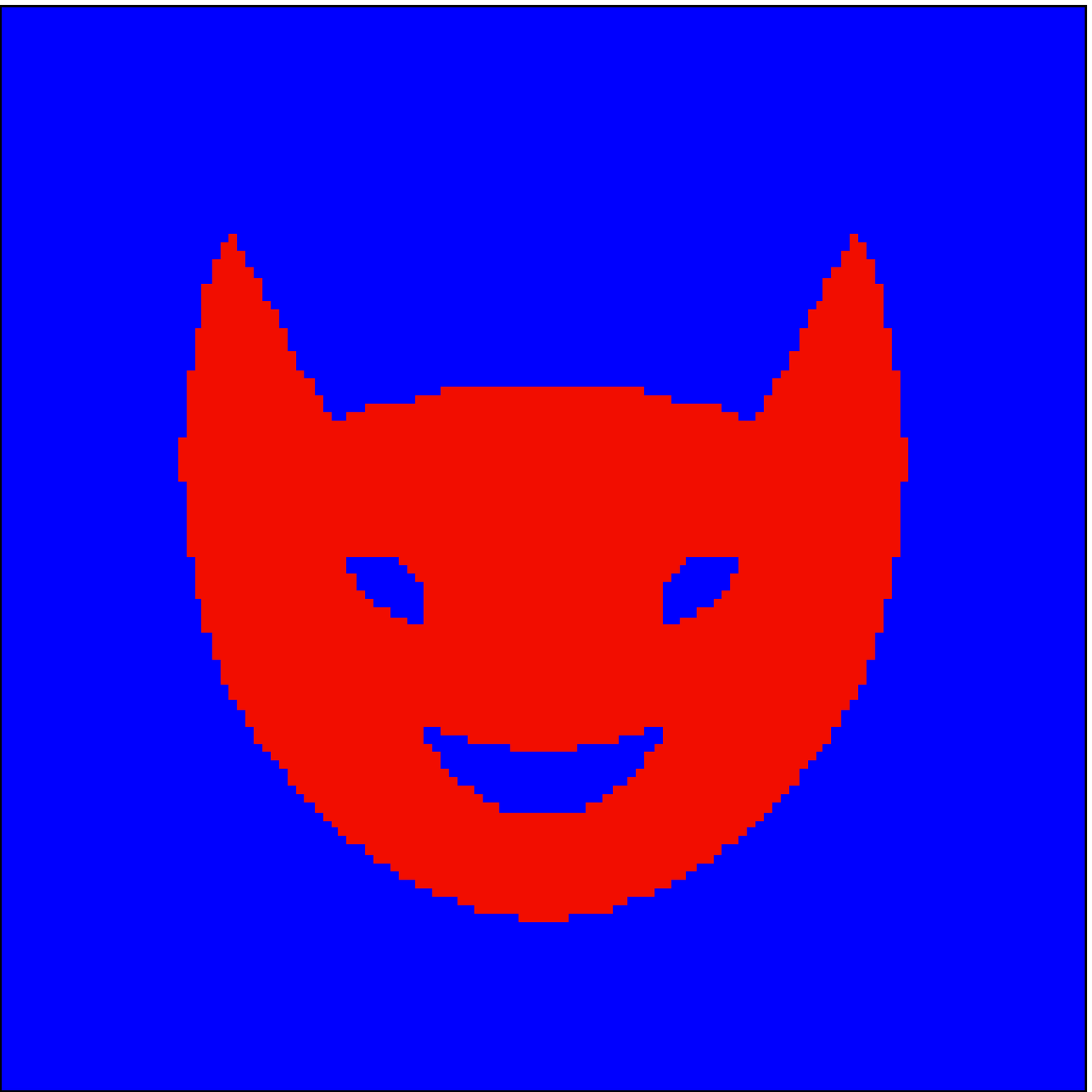}\\
\includegraphics[width=0.22\textwidth]{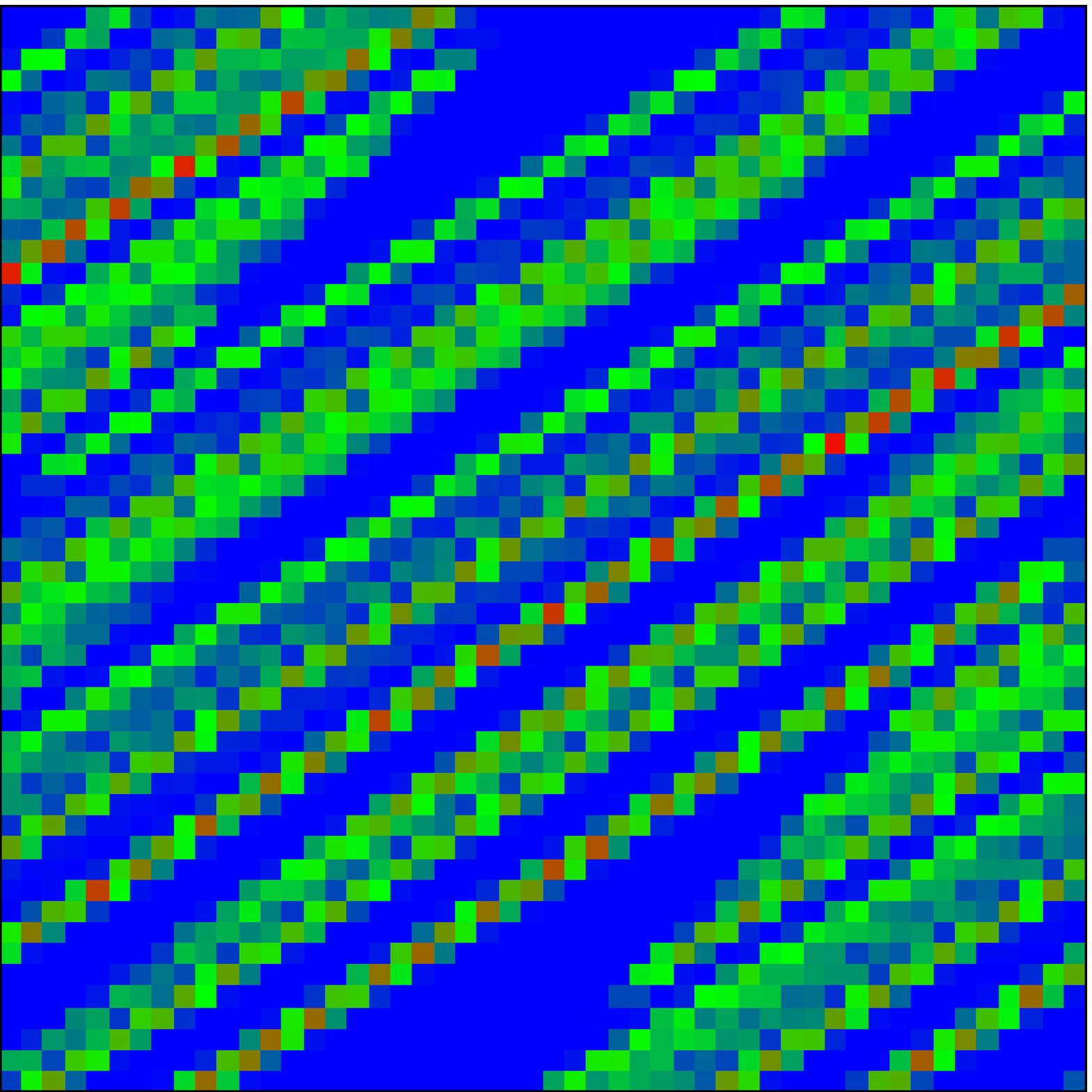}
\includegraphics[width=0.22\textwidth]{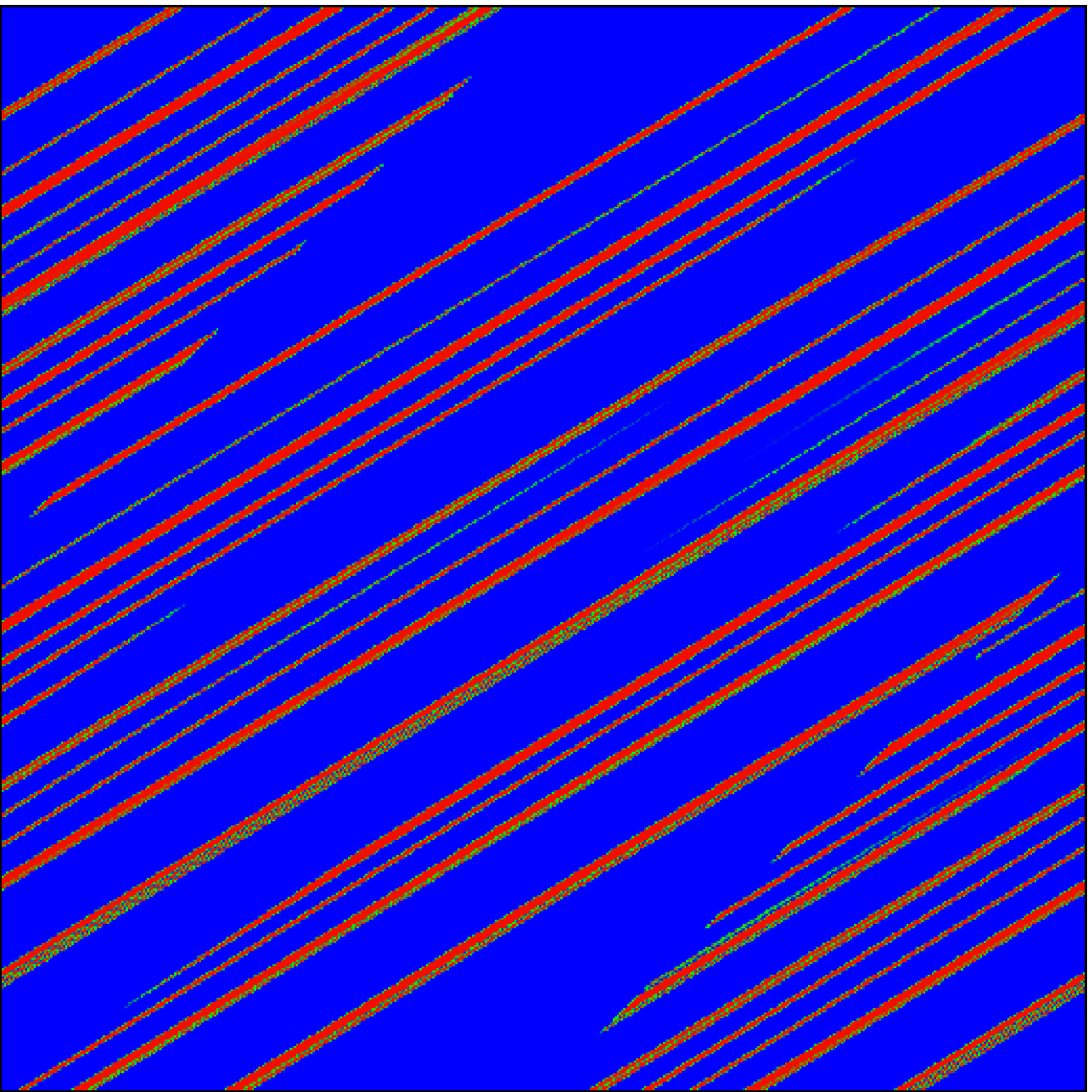}\\
\includegraphics[width=0.22\textwidth]{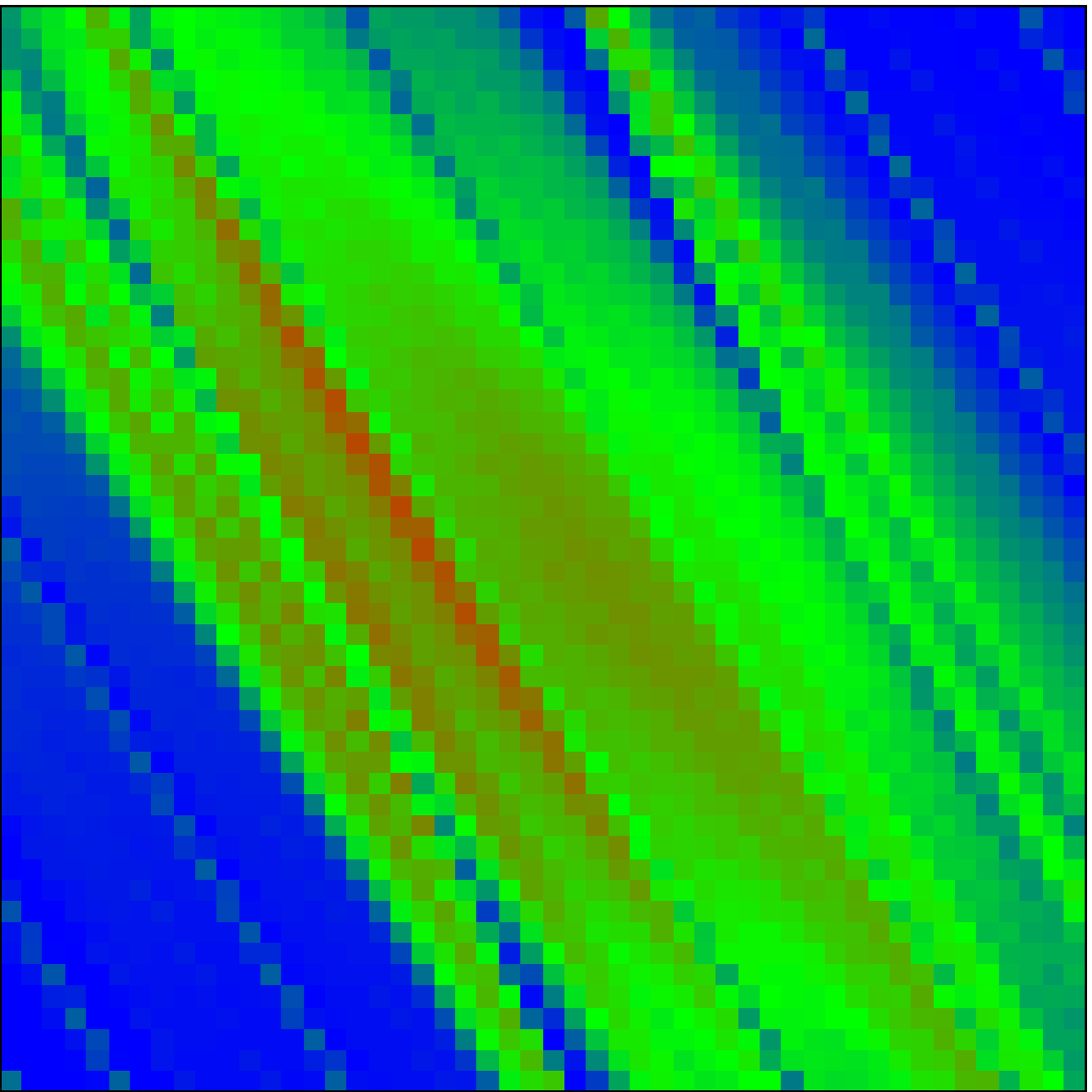}
\includegraphics[width=0.22\textwidth]{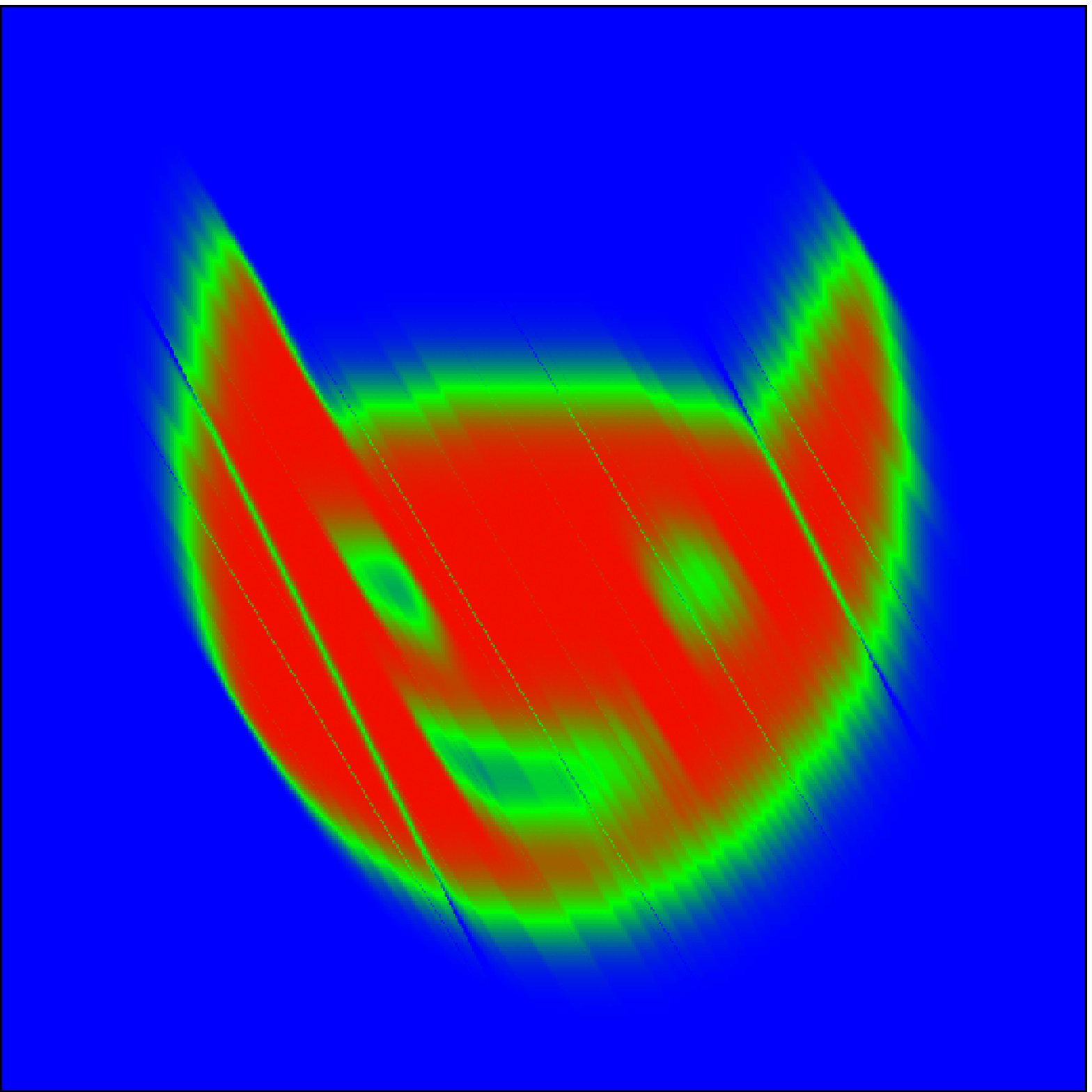}\\
\includegraphics[width=0.22\textwidth]{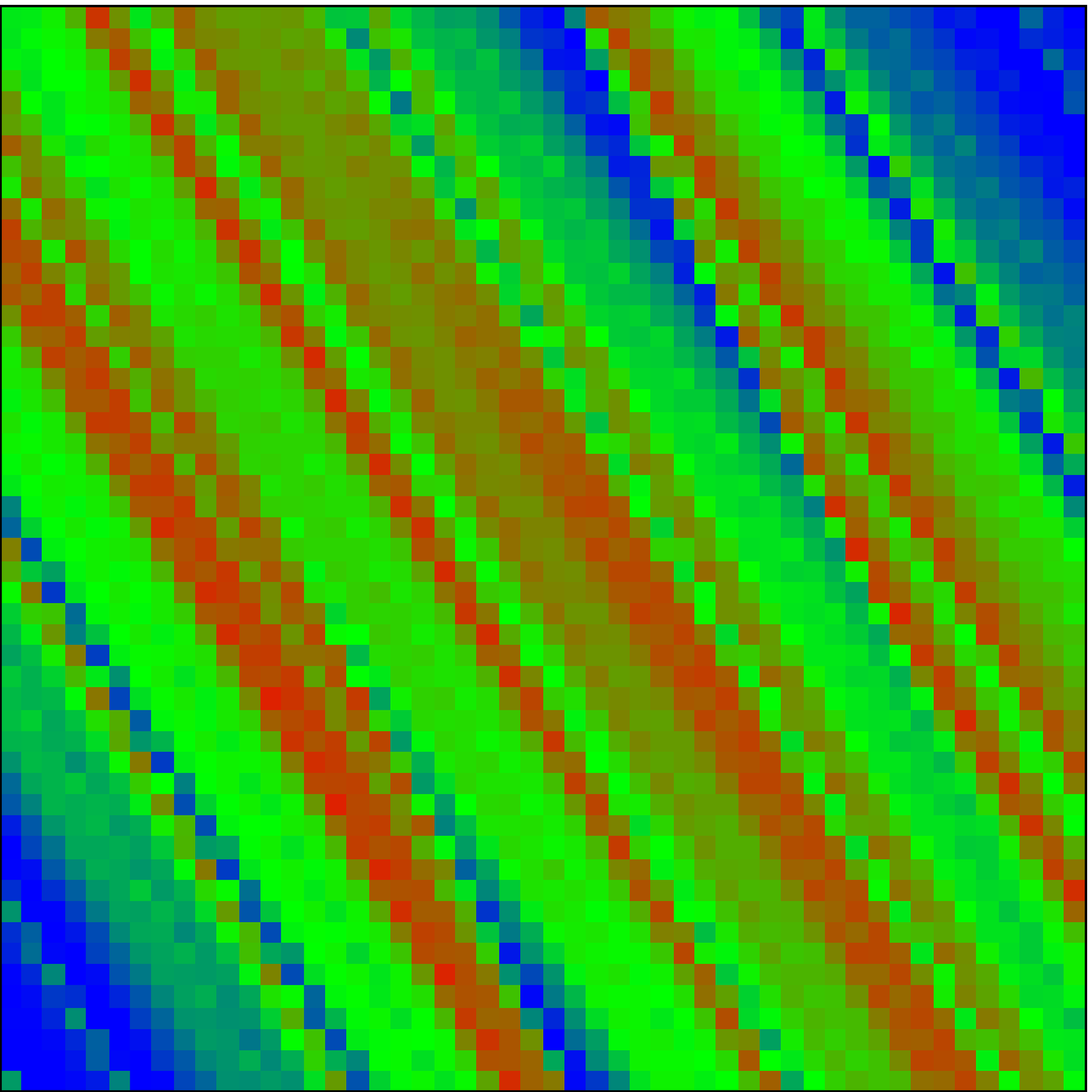}
\includegraphics[width=0.22\textwidth]{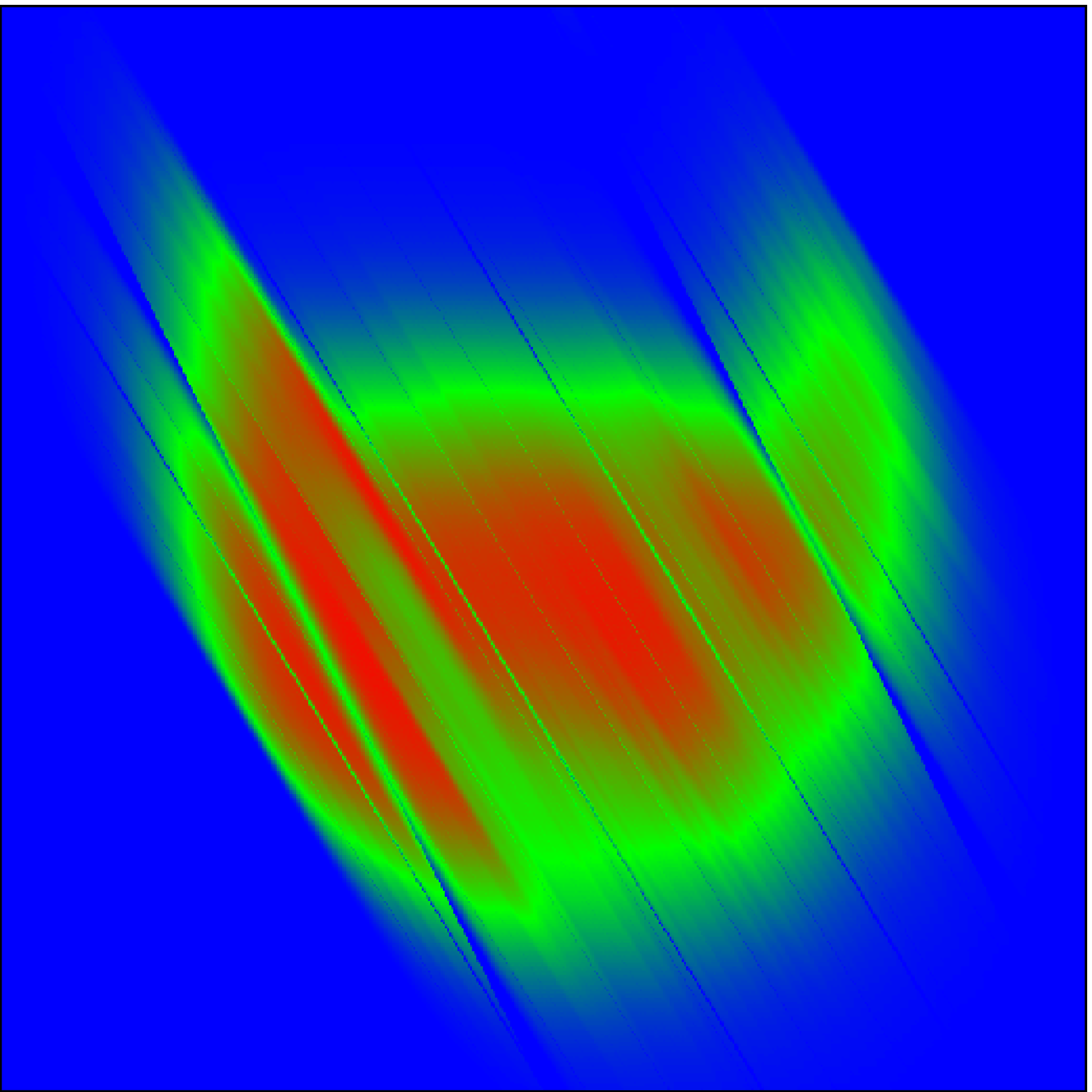}\\
\caption{(color online) Evolution of the Arnold cat image
generated by the Ulam matrix approximant $S$
with $N_x=51$ (left colum) and $N_x=751$ (right column)
at $L=8$, $N_p=L N_x$.
From top to bottom: initial image $t=0$;
image at the moment of time reversal $t=t_r=4$;
image at the moment of return $t=8$;
image at the return moment $t=10$ when the time reversal
is made after $t=t_r=5$ iterations.
Color is proportional to density with 
blue for zero and red for maximum;
 only the central initial section
$-0.5 \leq x < 0.5$, $-0.5 \leq x < 0.5$ is shown.
}\label{fig7}
\end{figure}

The strong effects of exponential instability on 
time reversal breaking are also well seen in Fig.~\ref{fig7}.
Here, the initial image of the Arnold cat 
cannot be recovered even if the time reversal is done
after only $t_r=4$
map iterations for the discretization level with
$N_x=51$.
Only for a much finer discretization level with
$N_x=751$ the initial image is approximately recovered
for $t_r=4$ but it degradates rapidly
already at time reversal performed after
$t_r=5$. This illustares the exponentially rapid breaking
of time reversal after the Ulam time $t_U$.

\begin{figure}
\includegraphics[width=0.44\textwidth]{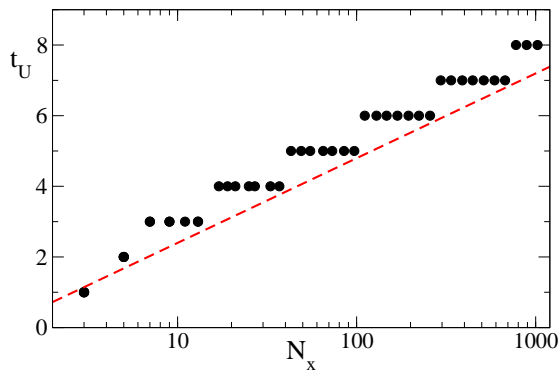}
\caption{(color online) Dependence of the Ulam time $t_{U}$
with anti-diffusion on the number of cells 
$N_x$ of the Ulam matrix approximant
at $L=8$ and $N_p=LN_x$. 
An initial state at momentum $p=0$ 
and homogeneous in $x$ is evolved 
up to time  $t=t_r=30$ 
when a time reversal 
operation is applied.  
The dashed line shows  the dependence $t_{U}=\ln{N_x}/h$
}\label{fig8}
\end{figure}

The dependence of 
the Ulam time $t_{U}$, characterized by the anti-diffusion
during time reversal process seen in Fig.~\ref{fig5},
on the discretization scale $N_x$ is shown in Fig.~\ref{fig8}.
We see that the results are well described by the dependence
\begin{equation}
\label{eq4}
 t_{U} = \ln N_x/h = |\ln \hbar_{eff}|/2h\;, \;\; \hbar_{eff}=1/N_x^2 \;\; ,
\end{equation}
where $\hbar_{eff}=1/N_x^2$ can be considered as an effective Planck
constant which gives the area of discretized cells.
In this form the Ulam time scale 
$t_{U}$ is similar to the Ehrenfest time $t_E$ 
\cite{chirikov1988} which
appears in the semiclassical limit of systems of quantum chaos.
In both cases the mechanism is related to the exponential growth of 
a wave packet of minimal size with time $\hbar_{eff} \exp(ht)$
due to which the packet size becomes comparable with the whole system size
after time $t_E$ or in our case after time $t_{U}$.

In spite of this similarity
we should note that in the quantum case the time reversal
is preserved under rather generic conditions
(see \cite{demon,qascat,dls1983} and Refs. therein).
In contrast to that
in the frame of the Ulam method, which also describes the linear
matrix evolution, the time reversal is 
broken after the Ulam time $t_U$.
The main reason of this difference is related to 
the fact that the Ulam matrix approximant
describes dynamics with eigenvalue modulus
smaller than unity while the quantum dynamics 
is unitary. This result can be also understood from the view point
of noise which have a size of discretized cells
and which also breaks time reversal.

\section{Discussion}
In this work we studied the properties of the Ulam matrix
approximant $S$, generated by the Ulam method, for the
Arnold cat map on a torus of a few integer sections $L$.
We show that the spectrum eigenvalues of $S$ converges to a   
limiting distribution in the limit of small cell discretization
and large matrix size $N$. The main part of this spectrum have
relaxation rates $\gamma$ with  approximate values of the Kolmogorov-Sinai
entropy in this system. There are also eigenvalues
with much smaller relaxation rate which is in a good agreement with the
statistical description by the Fokker-Planck equation.

The continuous model has the property of time reversal
but in the frame of the Ulam method the time
reversibility is broken on the Ulam time scale
$t_U$ which grows only logarithmically with 
the decrease of the cell size in the Ulam method.
Such a dependence has certain parallels with that one found 
for the Ehrenfest time scale in
systems of quantum chaos.
However, even if in both cases the evolution is described by
the linear matrix equations the quantum systems
preserves the property of time reversal in presence of weak perturbations, 
while for the Ulam method the time
reversal is broken after the Ulam time scale $t_U$.
Further studies are required for a better understanding of
relations between the spectrum of the Ulam matrix approximant,
chaos, diffusion, correlations decay and other statistical properties of
dynamical chaos in the Arnold cat map and other chaos systems.
Thus this simple model of Vladimir Arnold still keeps its
scientific  wonder.

{\bf Acknowledgments:} we thank Elena Chepelianskii 
for cat image drawing used here and in \cite{demon,qascat}.

\end{document}